\documentclass[12pt]{article}

\usepackage[cp1251]{inputenc}
\usepackage[english]{babel}
\usepackage{amssymb}
\usepackage{cite}
\usepackage{graphicx}

\newcommand{\reff}[1]{(\ref{#1})}
\newcommand{\refp}[1]{fig. \ref{#1}}

\begin{document}

\begin{center}

\textbf{THE FAILURE RISK ANALYSIS OF DIGITAL CIRCUITS}

\vskip\baselineskip
\textbf{\copyright\ 2013 A.N. Pchelintsev}

{\em Tambov State Technical University, ul. Sovetskaya 106,\\
Tambov, 392000 Russia\\
{\bf E-mail:} pchelintsev.an@yandex.ru}
\end{center}

\textbf{Abstract} -- To analyze the failure risk of asynchronous digital
circuits the time-parameter is introduced into the Boolean algebra replacing the
arithmetic operations by logical operations. There considered an example of
construction of signals passing through the logical elements, using the
described below mathematical apparatus.


\textbf{AMS}: 94C05, 06E25, 06E30.

\begin{center}
   1. INTRODUCTION
\end{center}

In Boolean algebra that is used for simulation of digital circuits, the
transition time (or inertia) of logical elements (eg, "and", "or") from one
state to another (eg, from 0 to 1) does not take into account. In cases when
the signal propagation time inside the element is sufficiently small the
switching delay can be ignored. But with increasing frequency of changes of the
input signals in real circuits the time influence starts to affect the signal
propagation inside its elements. Such delays may cause the unstable devices work
(i.e. there appeared a transitions, called failures in the signals after serial
passage through the nodes of the circuit unaccounted by the circuit model). Many
manufacturers of modern CPUs kept in secret how they struggle with failures
posed by the delays at frequencies of the order of GHz. In fact a common
conductor with many bends close to the board is converted into inductance in
this operation mode.

To analyze the most simple failure risk scheme there usually used the time
diagrams method \cite{1,2,3}, which has already become a classic. The signals at
each node are drawn strictly under each other: an artificial delay is produced
in transition from one state to another where it necessary, and then the output
signals are constructed according to Boolean representation. Given method is not
very good because it requires work with graphics that can make an error in the
received signals. We need to know whether there is a failure, and what will
it look like. That is why, in this article this procedure is transformed from
graphical representation to mathematical representation. At that we introduce a
time-parameter into Boolean algebra replacing the logic operations by
arithmetic. To simplify the analysis there considered asynchronous circuit, i.e.
uncontrolled by external (synchronizing or pulsing) signal digital circuit.

\begin{center}
   2. THE TRANSITION FROM THE LOGICAL REPRESENTATION TO ARITHMETIC
      REPRESENTATION OF BOOLEAN FUNCTIONS
\end{center}

Let us consider set of numbers $M=\{0;1\}$. It defined the operations of
negation, conjunction, disjunction, and their products (eg, implication,
disjunction, and other alternative). Let us express these logical operations
through the arithmetical on a set $M$:
\begin{equation}
   \label{perehod}
   \begin{array}{l}
      \overline{x}=1-x,\\
      x\wedge y=x\cdot y,\\
      x\vee y=x+y-x\cdot y.
   \end{array}
\end{equation}

Let us show the validity of law of De Morgan  $x\wedge y=\overline{\overline x
\vee\overline y}$:
$$
  \overline{\overline x\vee\overline y}=1-\overline x\vee\overline y=
  1-(1-x+1-y-(1-x)\cdot(1-y))=x\cdot y=x\wedge y.
$$

For expression \reff{perehod} let us add a rule
\begin{equation}
   \label{powerx}
   x^m=x
\end{equation}
for all natural values $m$, which  validity is obvious.

The expression for a Boolean function $f$, which is a function of input signals
of circuit can now be simplified by the laws of arithmetical operations and rule
\reff{powerx}. After some simplifications move back -- to Boolean
representation. At that the minimization process can be automated by using
symbolic calculations.

\begin{center}
   3. THE INJECTION OF TIME-PARAMETER INTO BOOLEAN ALGEBRA
\end{center}

As it is known the unit step function or Heaviside function is defined on the
area of real numbers and returns the number that belongs to the set $M$:
$$
  h(t)=
  \left\{
     \begin{array}{l}
        1, \: t\ge0,\\
        0, \: t<0.
     \end{array}
  \right.
$$

Let us denote the current time by $t$. Notice that function $h$ is also called
the turn off function. The following statement is obvious: any signal in the
logic circuit, comprising a transition from one logical state to another can be
represented as the sum of the difference of Heaviside functions, combined with
an appropriate argument.

For function $h$ there is the rule
\begin{equation}
   \label{hevi}
   \prod_{i=1}^n h(t-\tau_i)=h\left(t-\max_{i=\overline{1,n}}\tau_i\right),
\end{equation}
where $\tau_i$ -- time moment when there is a change in the signal. Let us add a
formula \reff{hevi} to \reff{perehod} and \reff{powerx}.

Now, knowing the analytical expression for the input signals of logical circuit,
there can be found a function form of the output signal.

\begin{center}
   4. DELAYS IN LOGIC CIRCUIT ELEMENTS
\end{center}

Is convenient to model a signal delay in the logic element as the difference
between the argument of the Heaviside function and the duration of the delay (so
as for existing logical elements, commonly, the delays along the front
(transition from 0 to 1) and recession (transition from 1 to 0) are
approximately equal). Thus, any real logical element of the circuit can be
modeled as a series connection of element of a pure delay \cite{4} for each
input and the ideal logic element (here the delay is equal to the duration of
delay). For example, the output signal equation of conjunctor delay $\tau$ on
the input takes the form:
$$
  y_c=f_1(t-\tau)\cdot f_2(t-\tau),
$$
where $f_1(t)$ and $f_2(t)$ -- functions that describe the corresponding input
signals.

\begin{center}
   5. THE SEARCH ALGORITHM OF FAILURE CONDITIONS
\end{center}

The proposed search algorithm of failure states is similar to time diagram
method; advantage of this method is that we work with graphical images signals,
and their analytical expressions (in this case it is possible to assess the
temporal characteristics of an analytical failure):
\begin{enumerate}
   \item Let the investigated scheme operates in accordance with a
   logical expression given by disjunctive-normal form;
   \item Defined by functions of input signals that represent transitions in the
   truth table, expressed in terms of the Heaviside function;
   \item Go along the path of signals in the logic circuit in order to find
   expressions for the output circuit, applying the rules of \reff{perehod} and
   \reff{hevi};
   \item If the resulting expression contains the difference between the
   Heaviside function, then we have a static failure, if there is a Heaviside
   function with delaying argument, then failure is dynamic.
\end{enumerate}

\begin{center}
   6. AN EXAMPLE OF LOGICAL SCHEME ANALYSIS
\end{center}

\begin{figure}[ht]
   \centering
   \includegraphics{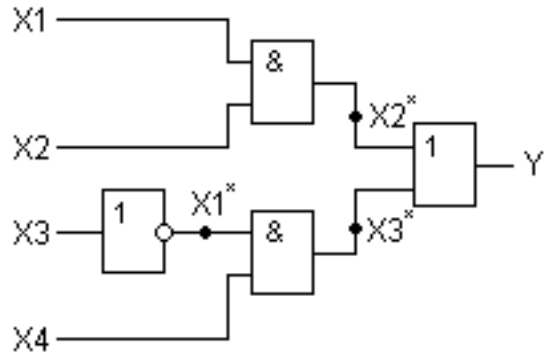}
   \caption{Circuit that realize the Boolean function
      $Y=X_1X_2\vee\overline{X_3}X_4$.}
   \label{picsh}
\end{figure}

\begin{figure}[ht]
   \centering
   \includegraphics{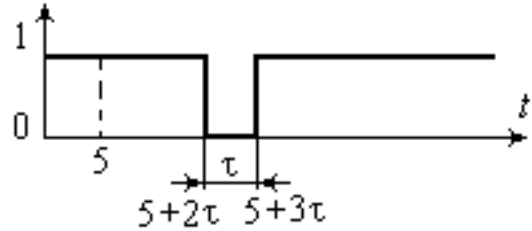}
   \caption{Signal diagram $Y(t)$.}
   \label{stat}
\end{figure}

Let us investigate the transition from the set 1111 to the set 1001
($15\rightarrow9$) of truth table for the circuit, that is shown at
\refp{picsh}.
Let us represent the input signals as follows (for simplicity, let us consider
the change of condition at the time moment, equal to 5 sec.).
$$
  \begin{array}{l}
     X_1=1,\\
     X_2(t)=1-h(t-5),\\
     X_3(t)=1-h(t-5),\\
     X_4=1.
  \end{array}
$$
Assume that all elements have the same delay, equal to $\tau$. Then
$$
  \begin{array}{l}
     X_1^*(t)=1-(1-h(t-5-\tau))=h(t-5-\tau),\medskip\\
     X_2^*(t)=1\cdot(1-h(t-5-\tau))=1-h(t-5-\tau),\medskip\\
     X_3^*(t)=X_1^*(t-\tau)\cdot1=h(t-5-2\tau),\medskip\\
     Y(t)=X_2^*(t-\tau)+X_3^*(t-\tau)-X_2^*(t-\tau)\cdot X_3^*(t-\tau)=
     1-h(t-5-2\tau)+\\+h(t-5-3\tau)-(1-h(t-5-2\tau))\cdot h(t-5-3\tau)=
     1-h(t-(5+2\tau))+\\+h(t-(5+2\tau))\cdot h(t-(5+3\tau))=1-h(t-(5+2\tau))+
     h(t-(5+3\tau)).
  \end{array}
$$
Thus, we have a static crash -- the difference of Heaviside functions is in the
resulting expression (\refp{stat}).

\begin{center}
   7. CONCLUSION
\end{center}

In this work there was considered the method of failure risk analysis of digital
circuits using an analytic representation of signals within the circuit. At that
logical operations had to by replaced by arithmetic operations. The transitions
from one signal state to another is described by the Heaviside function. The
advantage of the described modification of the time diagrams method is the
possibility of analytical analysis of the characteristics of failure.


\begin{thebibliography}{00}

\bibitem{1} Potemkin, I.S., {\em Functional units of digital automation}
(Russian), Jenergoatomizdat, Moscow, 1988.

\bibitem{2} Mulyarchik, S.G., {\em The integrated circuit design --
functional and logic level} (Russian), BSU, Minsk, 1983.

\bibitem{3} Pchelintsev, A.N., Kasyanov, A.N., {\em The analysis of
dangerous competitions in combinational digital schemes during automated
designing} (Russian), Transactions of the Tambov State Technical University,
Vol. 11, No. 2A (2005), pp. 368--371.

\bibitem{4} Besekerskij, V.A., Popov, E.P., {\em The theory of
automatic control systems} (Russian), Nauka, Moscow, 1975.

\end{thebibliography}
\end{document}